# Research on Heterogeneous Computation Resource Allocation based on Data-driven Method

Xirui Tang*
College of Computer Sciences
Northeastern University
Boston, MA, 02115, USA
* Corresponding e-mail address: tang.xir@northeastern.edu

Zeyu Wang
Indpendent Researcher
E-mail address: zeyu.wang.cs@ieee.org

Xiaowei Cai
Yale School of Management
Yale University
New Haven, US
E-mail address: chelsea.cai@yale.edu

Honghua Su
Olin Business School
Washington University in St. Louis
St. Louis, MO
E-mail address: ash.honghuasu@gmail.com

Changsong Wei
Digital Financial Information Technology Co.LTD
Chengdu, Sichuan, China
E-mail address: changsongwei88@gmail.com

*Abstract*—The rapid development of the mobile Internet and the Internet of Things is leading to a diversification of user devices and the emergence of new mobile applications on a regular basis. Such applications include those that are computationally intensive, such as pattern recognition, interactive gaming, virtual reality, and augmented reality. However, the computing and energy resources available on the user's equipment are limited, which presents a challenge in effectively supporting such demanding applications. In this work, we propose a heterogeneous computing resource allocation model based on a data-driven approach. The model first collects and analyzes historical workload data at scale, extracts key features, and builds a detailed data set. Then, a data-driven deep neural network is used to predict future resource requirements. Based on the prediction results, the model adopts a dynamic adjustment and optimization resource allocation strategy. This strategy not only fully considers the characteristics of different computing resources, but also accurately matches the requirements of various tasks, and realizes dynamic and flexible resource allocation, thereby greatly improving the overall performance and resource utilization of the system. Experimental results show that the proposed method is significantly better than the traditional resource allocation method in a variety of scenarios, demonstrating its excellent accuracy and adaptability.

*Keywords-heterogeneous computation, resource allocation, deep neural network, data-driven method.*

## I. INTRODUCTION

The rapid development of the Fifth Generation (5G) and the Internet of Things (IoT) has resulted in a significant increase in the number of user devices connected to the Internet, accompanied by a considerable rise in the volume of data and a proliferation of sophisticated mobile applications and services. Among these applications, those that are computationally intensive, such as pattern recognition, interactive games, and virtual/augmented reality, not only consume a considerable amount of computing resources and energy, but also require ultra-low response latency [1]. However, the constraints of hardware technology result in the computing resources, storage resources, and energy resources of mobile devices being limited to varying degrees. As a consequence, it is challenging to efficiently support the implementation of the aforementioned latency-sensitive and computing-intensive applications.

Users can offload computing tasks to the cloud server for processing, and use the sufficient computing and storage resources of the cloud server to meet the resource requirements of mobile applications. However, in the context of the Internet of Everything, massive data transmission and the emergence of computing-intensive applications have put forward more stringent requirements for limited uplink and downlink bandwidth and computing power [2]. Migrating a large amount of task data to an ECS puts a huge strain on the uplink and downlink bandwidth. Due to the distance between the cloud server and the user, factors such as network instability and routing uncertainty may lead to high transmission latency and poor reliability, which seriously affects the user experience. Therefore, cloud computing still has shortcomings in supporting latency-sensitive applications and achieving millisecond-level latency [3].

With the rapid development of mobile edge computing, there will be different specifications and different types of edge servers in the future network to meet the diverse computing task processing needs. Because the computing power of edge servers is often limited, different edge servers may need to work together to process computing tasks, which involves computing task offloading, multi-dimensional resource allocation, and computing result sharing [4]. However, the edge network is an open environment, and there may be malicious nodes in the network that may forge or tamper with compute offload requests, resource scheduling instructions, and computing results, resulting in the destruction of normal service processes.

In the field of computing, the method of obtaining higher computing power by increasing the number of CPU cores and clock frequency has encountered limitations in terms of energy consumption and heat dissipation. As the demand for computing power in a variety of fields continues to grow, an increasing number of computing systems are incorporating processors with diverse capabilities and physical architectures, including graphics processing units (GPUs), digital signal processors (DSPs), and field-programmable gate arrays (FPGAs), in order to achieve accelerated computing [5]. Such a system is referred to as a heterogeneous computing system. In order to achieve enhanced energy efficiency and computing performance, a diverse range of devices, including personal computers, smartphones, tablets, servers, and workstations, are evolving towards a heterogeneous computing model. Heterogeneous computing systems are capable of providing high-performance computing capabilities and are therefore of vital importance in the field of mobile communications, which are subject to strict requirements in terms of latency and energy consumption [6].

The heterogeneity of computing systems is reflected in the diversity of edge servers deployed in such systems. These servers offer a range of computing capabilities, enabling the provision of solutions tailored to diverse computing needs [7]. To illustrate, servers based on dedicated computing architectures can achieve accelerated computing by mining the parallelism and heterogeneity of specific types of applications. However, due to the specialised nature of the computing architecture, such servers may lack the capacity to handle other types of computing tasks.

## II. RELATED WORK

In heterogeneous computing systems, users anticipate that devices will be capable of running compute-intensive applications with minimal response latency and a sufficiently long battery life. Accordingly, the response delay of mobile applications and the battery life of mobile terminals serve as crucial metrics for gauging the quality of the user experience. However, the computing and energy resources of mobile terminals are constrained to varying degrees due to the limitations of hardware and technology [8]. The primary objective of heterogeneous computing systems is to address the issues of high latency and high energy consumption resulting from resource constraints when processing resource-intensive and latency-sensitive applications. Accordingly, the completion delay of tasks and the energy consumption of terminals have become significant metrics for evaluating the performance of heterogeneous computing systems.

Initially, Yang et al. [9] proposed a multi-user computing task block scheme for the first time. By assigning different users the offloading demand level determined by the delay demand (DLD), the multi-user computing offload is scheduled. This approach not only effectively coordinates and optimizes the computing resource requirements of multiple users, but also maximizes the overall performance and resource utilization efficiency of the system while meeting individual latency requirements. This scheme provides new ideas and methods for the promotion and optimization of heterogeneous computing systems in practical applications.

Additionally, Zhang et al. [10] used the small base station as a relay node with no computing power to assist users in computing offloading. With the goal of minimizing user energy consumption (MEC), the computing offloading decision and resource allocation are jointly optimized. This scheme not only reduces the energy consumption of user equipment, but also improves the efficiency of computing offloading, and provides new ideas and methods for the promotion and optimization of heterogeneous computing systems in practical applications.

Mao et al. [11] proposed a transmission power control and bandwidth resource allocation method based on Gauss-Seidel algorithm (GSA) with the goal of minimizing long-term average weighted energy consumption. This scheme not only reduces the energy consumption of user equipment, but also improves the efficiency of computing offloading, and provides new ideas and methods for the promotion and optimization of heterogeneous computing systems in practical applications.

Further, Lyu et al. [12] defined the weighted sum of task completion delay and user energy consumption as the system utility function, and maximized the system utility value by jointly optimizing the computational offloading decision, communication and computing resource allocation. This method can not only effectively reduce the energy consumption of user equipment, but also significantly improve the efficiency of computing offloading and the overall performance of the system, which provides new ideas and methods for the promotion and optimization of heterogeneous computing systems in practical applications.

## III. METHODOLOGIES

### A. Feature extraction and resource prediction

In the feature extraction phase, we transform the raw data we collect into features that the model can process. The purpose of feature extraction is to identify key variables related to resource requirements. Common characteristics include CPU utilization, memory usage, network bandwidth, task execution time, resource type, and time-dependent characteristics. To improve the accuracy of the prediction, we also calculate some derived features. At time t moment, the raw data collected is $d_t = [d_t^1, d_t^2, \ldots, d_t^M]$, where $d_t^i$ represents the i-th original feature, and $M$ is the number of original features. Through feature engineering, we construct an eigenvector $x_t = [x_t^1, x_t^2, \ldots, x_t^N]$, where $x_t^j$ represents the j-th extracted feature, and $N$ is the number of extracted features. We calculate the moving average derivative feature, which is expressed as Equation 1, where $k$ is the size of the moving window.

$$x_t^{N+1} = \frac{1}{k} \sum_{i=0}^{k-1} CPU_{t-i} \qquad (1)$$

After the feature extraction is complete, we use a deep neural network for resource demand forecasting. Suppose we want to predict the resource demand $\hat{r}_{t+1}$ at time $t+1$ in the future, and the goal of the forecasting model is to minimize the prediction error. The input of the predictive model is the eigenvector $x_t$ and the output is the resource demand vector $\hat{r}_{t+1}$, which has the same dimensions as the number of resource types. We use deep neural network for modelling, and its structure consists of an input layer, multiple hidden layers, and an output layer. The general framework is illustrated as Figure 1.

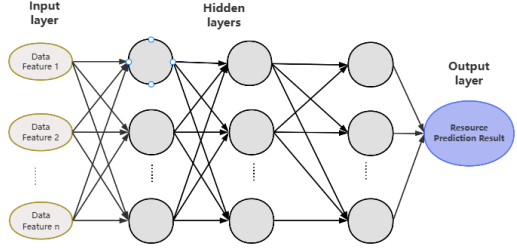

Figure 1. Framework of Proposed Feature Prediction

The loss function is used to measure the error between the predicted value and the actual value, and commonly used loss functions include the mean square error, which is expressed as Equation 2, where $T$ is the sample size and $r_{t+1}$ is the actual resource requirement.

$$L = \frac{1}{T}\sum_{t=1}^{T}(r_{t+1}, \hat{r}_{t+1})^2 \quad (2)$$

By optimizing the parameters of the proposed model $\theta = \{W_l, b_l\}_{l=1}^{L+1}$ by the backpropagation algorithm, we can minimize the loss function and thus improve the prediction accuracy.

### B. Dynamic resource allocation optimization

After resource demand forecasting, dynamic resource allocation optimization is next. The goal of this part is to maximize the system utility function by optimizing computational offloading decisions, communication, and computational resource allocation. The system utility function is defined as the weighted sum of the task completion delay and the user's energy consumption. The system utility function U is expressed as Equation 3.

$$U = \sum_{i=1}^{N}(\alpha_i TET_i + \beta_i Energy_i) \quad (3)$$

Where $N$ is the number of tasks, $\alpha_i$ and $\beta_i$ are the time weights and energy consumption weights of the ith task, $TET$ is the task execution time, and $Energy$ is the energy consumption of users.

The goal of joint optimization is to maximize system utility by taking into account computational offloading decisions, communication, and computing resource allocation simultaneously. To achieve this, we use a hybrid integer linear programming approach to solve complex multi-objective optimization problems. In our resource allocation problem, the variables include the calculation of the offload decision u, the transmission power p, and the bandwidth allocation b. We define the joint optimization problem as Equation 4.

$$\min_{u,p,b} \sum_{i=1}^{N}(\alpha_i \cdot (\sum_{j=1}^{M} u_{ij} \cdot TET_{ij} + (1 - \sum_{j=1}^{M} u_{ij}) \cdot TET_{i,local}) +$$
$$\beta_i \cdot (\sum_{j=1}^{M} u_{ij} \cdot Energy_{ij} + (1 - \sum_{j=1}^{M} u_{ij}) \cdot Energy_{i,local})) \quad (4)$$

Where $M$ is the number of edge servers, $u_{ij}$ indicates whether the i-th task is offloaded to the j-th edge server, $TET_{ij}$ and $Energy_{ij}$ are the execution time and energy consumption of task $i$ on the edge server $j$, $TET_{i,local}$ and $Energy_{i,local}$ are the time and energy consumption of task $i$ to be executed locally, respectively.

## IV. EXPERIMENTS

### A. Experimental setups

We set the number of user unloading tasks to be processed by the server is 15, and the uplink transmission rate of users based on the unit bandwidth resources is evenly distributed at $1 \times 10^6, 2 \times 10^6$ bps, and the number of bandwidth resources is 50. The server configuration consists of a 2.4 GHz 4-core CPU processor (each delivering 9 Gcycles/s) and a 1038MHz 768-core GPU processor (each delivering 100 Gcycles/s) with 5 CPUs and 5 GPU processors, respectively. Suppose that the occurrence probability of a special task is 0.7, the average data volume of a common task is 420 KB, the average CPU computation required is 1 Gcycles, the GPU utilization efficiency of a special task is 0.2, and the task latency sensitivity is evenly distributed at 0,1. The weights of user task processing efficiency and block generation efficiency are 0.8 and 0.2, respectively, and the minimum CPU and GPU computing resources required to process the task are 0.1 Gcycles/s and 1 Gcycles/s, respectively. The goal of optimization is to maximize the system utility function by jointly optimizing the calculation of offload decisions, transmission power, and bandwidth allocation.

### A. Experimental analysis

Task execution time (TET) refers to the total time from when a task is unloaded to the server to start processing, to when the task is processed and the result is returned, and is an important indicator to measure the response speed and performance of the system. Figure 2 compares the task execution comparison results.

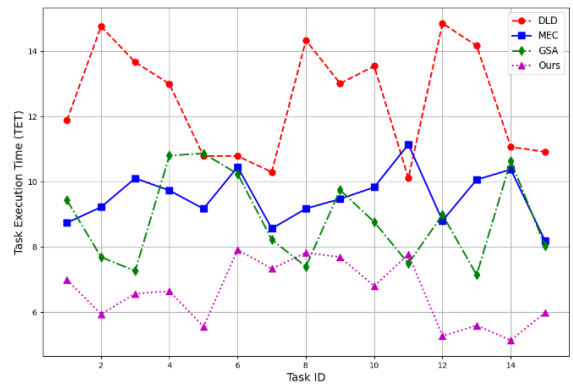

Figure 2. Task Execution Time Comparison

In Figure 2, we compare the performance of the four methods (DLD, MEC, GSA, Ours) in terms of task completion delay. The results show that our method (Ours) exhibits a lower task completion delay in all tasks, which is significantly better than other methods. This shows that our approach has advantages in optimizing task processing efficiency and reducing system response time.

User energy consumption refers to the energy consumed in the process from unloading to completion, which is an important indicator to measure the energy-saving performance of the system. Energy consumption is affected by factors such as task completion latency, processor power consumption, and the type of computing resources. Figure 3 experiments compare the user energy consumption of the four methods (DLD, MEC, GSA, Ours) and show that our method (Ours) exhibits lower energy consumption in all tasks, significantly better than the other methods. This shows that our approach has advantages in optimizing energy consumption, helping to extend the battery life of mobile devices and IoT devices, and improve the energy efficiency of the system.

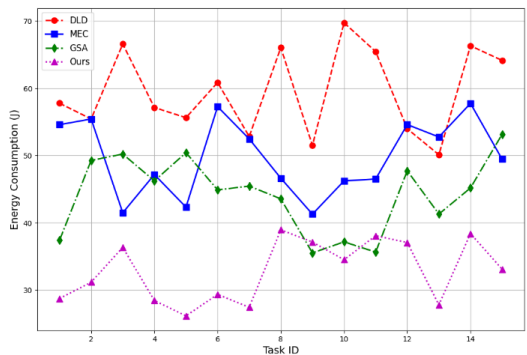

Figure 3. User Energy Consumption Comparison

Task Completion Rate refers to the ratio of the number of tasks successfully completed in a given time to the total number of tasks, and is an important indicator to measure the efficiency of the system in processing tasks. It is affected by factors such as task completion latency, resource allocation policies, and system load.

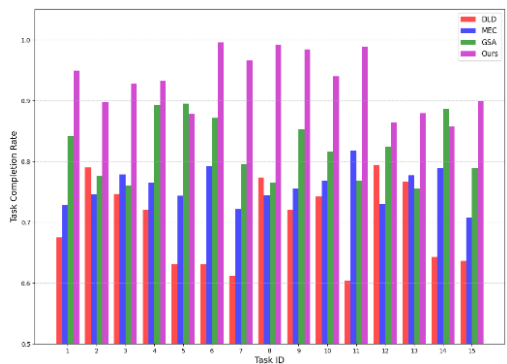

Figure 4. Task Completion Rate Comparison

Figure 4 compares the task completion rates of the four methods (DLD, MEC, GSA, Ours), and the results show that our method (Ours) shows that our method (Ours) shows a higher task completion rate in all tasks, which is significantly better than the other methods, indicating that our method has significant advantages in improving the task processing efficiency of the system.

## V. Conclusions

In conclusion, the heterogeneous computing resource allocation model based on the data-driven approach performs well in key performance indicators such as task completion delay, user energy consumption, and task completion rate by predicting future resource requirements and dynamically optimizing them. The experimental results show that our method (Ours) is significantly better than other methods (DLD, MEC, GSA), and shows lower task completion delay, user energy consumption and higher task completion rate in all tasks. The conclusion is that the data-driven approach can effectively improve the overall performance and resource utilization of the system, and provides an efficient and reliable solution for resource allocation in heterogeneous computing environments. As for future improvements, with the growth of computing demand and the increasing variety of computing resources, the data-driven resource allocation model will have a broader application prospect. Further research can explore more complex prediction algorithms and optimization strategies to cope with more diverse tasks and more complex computing environments.

REFERENCES

[1] Mohajer, Amin, et al. "Heterogeneous computational resource allocation for NOMA: Toward green mobile edge-computing systems." IEEE Transactions on Services Computing 16.2 (2022): 1225-1238.

[2] Ding, Changfeng, et al. "Joint MIMO precoding and computation resource allocation for dual-function radar and communication systems with mobile edge computing." IEEE Journal on Selected Areas in Communications 40.7 (2022): 2085-2102.

[3] Xu, Haitao, et al. "Edge computing resource allocation for unmanned aerial vehicle assisted mobile network with blockchain applications." IEEE Transactions on Wireless Communications 20.5 (2021): 3107-3121.

[4] Tan, Lin, et al. "Energy-efficient joint task offloading and resource allocation in OFDMA-based collaborative edge computing." IEEE Transactions on Wireless Communications 21.3 (2021): 1960-1972.

[5] Du, Jun, et al. "SDN-based resource allocation in edge and cloud computing systems: An evolutionary Stackelberg differential game approach." IEEE/ACM Transactions on Networking 30.4 (2022): 1613-1628.

[6] Cao, Bin, et al. "Resource allocation in 5G IoV architecture based on SDN and fog-cloud computing." IEEE transactions on intelligent transportation systems 22.6 (2021): 3832-3840.

[7] Li, Xianwei, et al. "A cooperative resource allocation model for IoT applications in mobile edge computing." Computer Communications 173 (2021): 183-191.

[8] Wei, Wenting, et al. "Multi-objective optimization for resource allocation in vehicular cloud computing networks." IEEE Transactions on Intelligent Transportation Systems 23.12 (2021): 25536-25545.

[9] Yang, Lei, et al. "Multi-user computation partitioning for latency sensitive mobile cloud applications." IEEE Transactions on Computers 64.8 (2014): 2253-2266.

[10] Zhang, Ke, et al. "Energy-efficient offloading for mobile edge computing in 5G heterogeneous networks." IEEE access 4 (2016): 5896-5907.

[11] Mao, Yuyi, et al. "Stochastic joint radio and computational resource management for multi-user mobile-edge computing systems." IEEE transactions on wireless communications 16.9 (2017): 5994-6009.

[12] Lyu, Xinchen, et al. "Multiuser joint task offloading and resource optimization in proximate clouds." IEEE Transactions on Vehicular Technology 66.4 (2016): 3435-3447.